\documentclass[aps,showpacs,twocolumn,epsf,dvips]{revtex4}
\usepackage{feynmp}
\usepackage{amssymb}
\usepackage{epsfig}
\usepackage{graphicx}
\usepackage{subfigure}
\usepackage{color}
\begin{document}

\title{Unconventional Fermi surface in two-dimensional systems of Dirac fermions}
\author{Jing-Rong Wang and Guo-Zhu Liu \\
{\small {\it Department of Modern Physics, University of Science
and Technology of China, Hefei, Anhui, 230026, P.R. China }}}

\begin{abstract}
At the low energy regime, the decay rate of two-dimensional massless
Dirac fermions due to interactions can be written as
$\mathrm{Im}\Sigma(\omega) \propto |\omega|^{x}$ at zero temperature.
We find that the fermion system has: I) no sharp Fermi surface and
no well-defined quasiparticle peak for $0<x<\frac{1}{2}$; II) a
sharp Fermi surface but no well-defined quasiparticle peak for
$\frac{1}{2}\leq x\leq 1$; III) both sharp Fermi surface and
well-defined quasiparticle peak for $x>1$. In the presence of
long-range gauge/Coulomb interaction or certain massless boson mode,
the system exhibits unusual behavior belonging to class II.
\end{abstract}


\maketitle


Landau's Fermi liquid (FL) theory is the standard model of quantum
many-body physics \cite{Abrikosov}. During the past five decades,
this theory has been applied successfully to a wide range of
condensed matter systems. In any normal FL, there are always sharp
Fermi surface and well-defined quasiparticle peak at the Fermi
energy. Generally, the fermion decay rate can be written as
$\mathrm{Im}\Sigma(\omega) \propto |\omega|^{x}$ for small energy
$\omega$ at zero temperature. The quasiparticles are well-defined
only when the decay rate vanishes faster than $\omega$ does upon
approaching the Fermi surface. Hence, an interacting fermion system
is a normal FL for $x>1$ and a non-FL for $x\leq 1$. One powerful
quantity to characterize the FL is the renormalization factor $Z$,
also called the quasiparticle residue, which is defined by the real
part of the retarded fermion self-energy $\mathrm{Re}\Sigma(\omega)$
as
\begin{equation}
Z = \left.\left(1-\frac{\partial \mathrm{Re}\Sigma(\omega)}{\partial
\omega}\right|_{\omega=0}\right)^{-1}.
\end{equation}
It is easy to check that $Z=0$ for $0<x \leq 1$ but $0<Z< 1$ for
$x>1$. Therefore, the residue $Z$ vanishes in a normal FL and has a
finite value in a non-FL. Physically, $Z$ measures the discontinuity
of fermion momentum occupation number $n(\mathbf{k})$ at Fermi
energy, which signals the presence of a sharp Fermi surface.

According to the conventional wisdom, the finiteness of residue $Z$
is sufficient to guarantee the presence of a sharp Fermi surface and
well-defined quasiparticle peak. However, the opposite reasoning may
not be correct. In particular, the presence of sharp Fermi surface
does not necessarily require that $Z$ is finite. Indeed, there is an
interesting possibility that a system may have a sharp Fermi surface
even though $Z=0$. Recently, Senthil \cite{Senthil} studied the
unusual properties of the quantum critical point of a Mott insulator
to metal phase transition. In his scenario, although $Z=0$ at the
critical point, there can be a sharp Fermi surface, dubbed as
critical Fermi surface, that is characterized by a kink singularity,
i.e., the discontinuity in the derivative of the fermion occupation
number $n(\mathbf{k})$ at the Fermi energy \cite{Senthil}. On the
experimental side, it has long been known that there is Fermi
surface (or Fermi arc) in the normal state of underdoped high
temperature superconductors despite the absence of well-defined
Landau quasiparticles \cite{Shenzx, Norman, Taillefer}.

In this paper, we propose that this possibility can be realized in
some planar interacting systems composed of massless Dirac fermions.
In recent years, the Dirac fermions with linear dispersion have been
investigated extensively because they are the low-energy excitations
of many strongly correlated systems, including $d$-wave high
temperature superconductor \cite{Shenzx}, quantum Hall system
\cite{Ludwig}, graphene \cite{CastroNeto}, and iron-based
superconductor \cite{Hasan}. In different physical systems, these
fermions may experience gauge, Coulomb, or other kinds of
interactions. We show that the Dirac fermion system has a sharply
defined Fermi surface but no quasiparticle peak when the exponent
$x$ of decay rate satisfies $\frac{1}{2}\leq x \leq 1$. In contrast
to the scenario of Senthil, the Fermi surface in our case is defined
by the divergence of the derivative of $n(\mathbf{k})$ at the Fermi
energy, rather than a kink singularity. When the Dirac fermions
couple to massless boson modes, such as gauge field, Coulomb
potential, or certain order parameter fluctuation, they exhibit such
unconventional behavior. For $0 < x< \frac{1}{2}$, the fermion
system has no sharp Fermi surface and no quasiparticle peak, which
loses any similarity to a free fermion gas. For $x > 1$, there are
both Fermi surface and quasiparticle peak, so the system is just a
normal FL.

We begin with the following free action
\begin{equation}
\mathcal{L}_0 = \sum_{i=1}^{N} \Psi_{i}^\dag\left(\partial_{\tau} -
iv_F\mathbf{\sigma}\cdot \mathbf{\partial}\right)\Psi_{i},
\end{equation}
where spinor field $\Psi$ describes massless Dirac fermions and
$v_F$ is the fermion velocity. The coupling of Dirac fermions with
various singular boson modes can result in unusual behaviors.

We first consider the interaction of Dirac fermions with an abelian
gauge field, which corresponds to three-dimensional quantum
electrodynamics (QED$_3$). After proper modifications, this model
can be used to study the low-energy properties of $d$-wave high
temperature superconductor \cite{Affleck, Liu} and spin liquid state
\cite{Ran07}. Recently, we studied the decay rate of massless Dirac
fermions due to gauge interaction. At zero chemical potential
$\mu=0$, the fermion decay rate is always divergent within
perturbation expansion method. We calculated the decay rate by the
self-consistent Eliashberg equation approach and found
\cite{WangLiu} that it is of the form $\mathrm{Im}\Sigma(\omega)
\propto |\omega|^{1/2}$ at zero temperature. At finite $\mu$, the
longitudinal component of gauge field becomes massive due to static
screening, but the transverse component remains massless. In this
case, the decay rate is free of divergence and depends on energy as
$\mathrm{Im}\Sigma(\omega) \propto \mu^{-1/3}|\omega|^{2/3}$ at zero
temperature \cite{WangLiu2}. Obviously, the chemical potential only
changes the coefficient of fermion decay rate. When applied to
cuprate superconductor, the gauge field may couple to an additional
scalar field \cite{Leereview}. In the superconducting ground state,
the gauge field acquires a finite mass via the Anderson-Higgs
mechanism. It is easy to obtain that, $\mathrm{Im}\Sigma(\omega)
\propto |\omega|^{3}$, which has the same energy dependence as the
decay rate yielded by contact four-fermion interaction
\cite{Yashenkin}.

In the context of graphene, the massless Dirac fermions experience
the Coulomb interaction, which is unscreened due to the vanishing
density of states at Fermi energy \cite{CastroNeto}. From previously
analysis \cite{Gonzalez, WangLiu}, we know that the decay rate
caused by Coulomb interaction is $\mathrm{Im}\Sigma(\omega) \propto
|\omega|$, which is clearly marginal FL behavior \cite{Varma}.

In the vicinity of continuous quantum phase transition, the massless
Dirac fermions interact with the strong fluctuation of order
parameter \cite{Vojta03}. Such interaction can be described by a
Yukawa coupling between a spinor field and a scalar field. In
particular, at the critical point between two supercondcuting phases
in $d$-wave cuprate superconductor \cite{Vojta} , the fermion decay
rate due to massless order parameter fluctuation is
$\mathrm{Im}\Sigma(\omega) \propto |\omega|$.

In general, after including the self-energy correction due to
various interactions, the full retarded Green function of Dirac
fermion is
\begin{equation}
G(\omega,\mathbf{k}) = \frac{1}{\omega-\xi_{\mathbf{k}}
-\mathrm{Re}\Sigma(\omega) - i\mathrm{Im}\Sigma(\omega)},
\end{equation}
where $\xi_{\mathbf{k}}=\varepsilon_{\mathbf{k}}-\mu$ with $\mu$
being the chemical potential and $\varepsilon_{k} = v_F|\mathbf{k}|$
energy of the Dirac fermion. To simplify the problem, here we assume
that the self-energy depends only on energy $\omega$. The spectral
function, defined as $A(\omega,\mathbf{k}) =
-\frac{1}{\pi}\mathrm{Im}G(\omega,\mathbf{k})$, has the form
\begin{eqnarray}
A(\omega,\mathbf{k}) = -\frac{1}{\pi}\frac{\mathrm{Im}
\Sigma(\omega)}{\left[\omega - \varepsilon_{\mathbf{k}}
-\mathrm{Re}\Sigma(\omega)\right]^2 + \left[\mathrm{Im}
\Sigma(\omega)\right]^2}.
\end{eqnarray}
The fermion momentum occupation number is given by
\begin{equation}
n(\mathbf{k}) = \int_{-\infty}^{0}d\omega
A\left(\omega,\mathbf{k}\right).
\end{equation}
In order to make a general analysis, we write the zero-temperature
decay rate of Dirac fermions as
\begin{equation}
\mathrm{Im}\Sigma(\omega) = C|\omega|^{x},
\end{equation}
where $x$ is a tuning parameter and $C$ a negative constant. The
real part of self-energy can be obtained from the Kramers-Kronig
relation
\begin{equation}
\mathrm{Re}\Sigma(\omega) = \frac{1}{\pi}
P\int_{-\infty}^{+\infty}d\omega'\frac{\mathrm{Im}
\Sigma(\omega')}{\omega'-\omega}. \label{eqn:KKRelation1}
\end{equation}

We first consider the region $0<x<1$. In such region, the real
self-energy function is found to be
\begin{equation}
\mathrm{Re}\Sigma(\omega) = C\mathrm{sgn}(\omega)|\omega|^{x}I(x),
\end{equation}
where $I(x)$ is
\begin{eqnarray}
I(x) &=& \frac{1}{\pi x}\lim_{\delta\rightarrow0}
[\int_{0}^{+\infty} dy \frac{1}{y^{\frac{1}{x}}+1} +
\int_{0}^{1-\delta} dy \frac{1}{y^{\frac{1}{x}}-1} \nonumber \\
&& + \int_{1+\delta}^{+\infty} dy \frac{1}{y^{\frac{1}{x}}-1}].
\end{eqnarray}
When $0<x<1$, the function $I(x)$ has a finite magnitude and hence
the real part of retarded self-energy $\mathrm{Re}\Sigma(\omega)$
has the same energy dependence as $\mathrm{Im}\Sigma(\omega)$. In
particular, we have $I(1/2)=1$ and $I(2/3)=\sqrt{3}$.

When $x\geq1$, there will be divergence when
Eq.(\ref{eqn:KKRelation1}) is used to calculate
$\mathrm{Re}\Sigma(\omega)$. To get a finite result, we should
introduce a cutoff energy $\omega_{c}$ and write
\begin{equation}
\mathrm{Re}\Sigma\left(\omega\right) = \frac{1}{\pi}
P\int_{-\omega_{c}}^{+\omega_{c}}d\omega'\frac{\mathrm{Im}
\Sigma\left(\omega'\right)}{\omega'-\omega}. \label{eqn:KKRelation2}
\end{equation}
For $x=1$, we have
\begin{eqnarray}
\mathrm{Re}\Sigma(\omega) = C\mathrm{sgn}(\omega)|\omega|
\frac{1}{\pi}\left[\ln\left(\frac{\omega_c+|\omega|}{|\omega|}\right)
+\ln\left(\frac{\omega_c - |\omega|}{|\omega|}\right)\right], \nonumber
\end{eqnarray}
which in the low-energy regime reduces to
\begin{equation}
\mathrm{Re}\Sigma(\omega) \approx
C\frac{2}{\pi}\omega\ln\left(\frac{\omega_c}{|\omega|}\right).
\end{equation}
This corresponds to the marginal FL behavior. For $x>1$, the real
part of self-energy at low energy is
\begin{eqnarray}
\mathrm{Re}\Sigma(\omega) \approx
C\frac{2}{\pi}\frac{|\omega_c|^{x-1}}{x-1}\omega.
\end{eqnarray}

\begin{figure}[h]
   \subfigure{
    \includegraphics[width=1.4in]{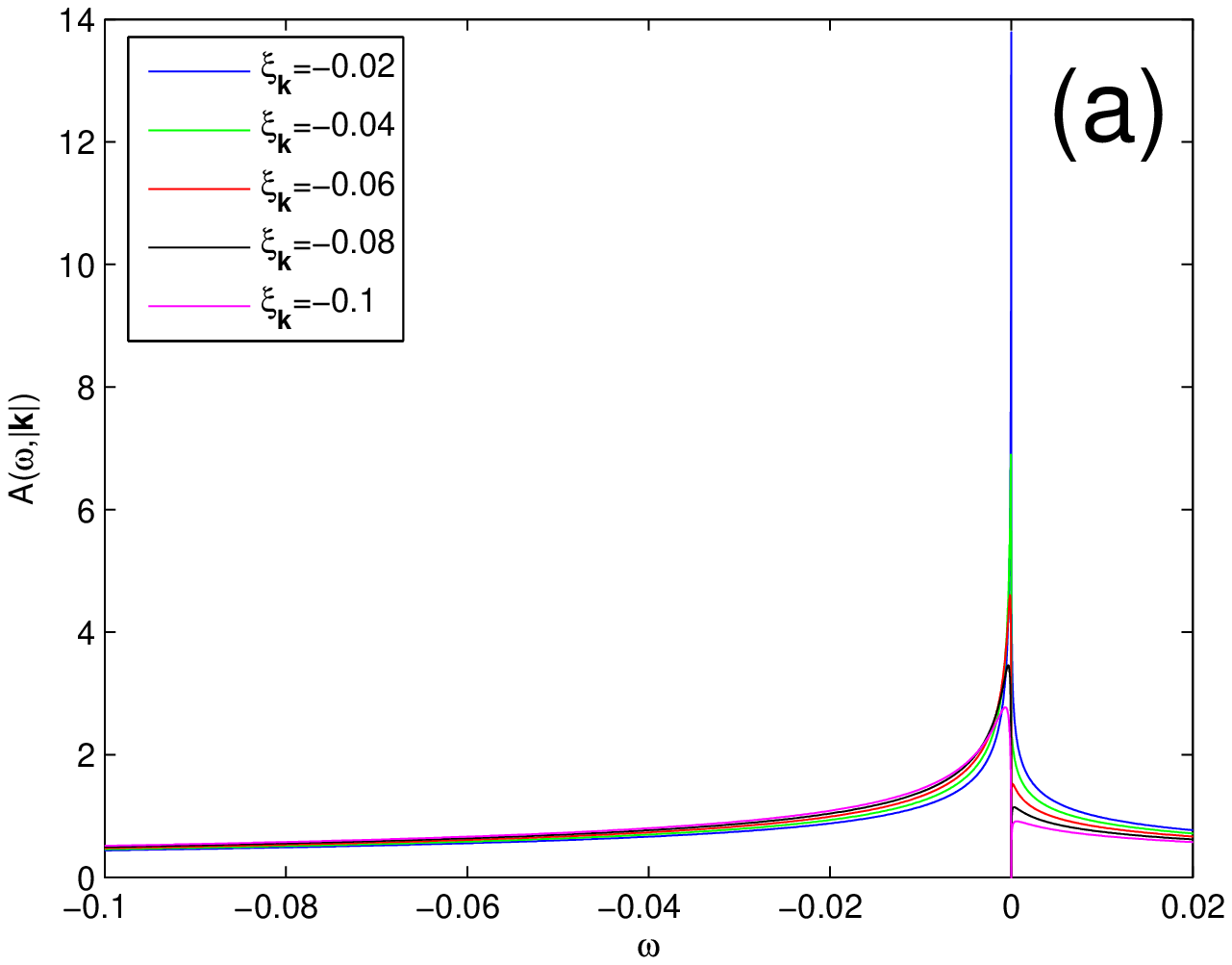}}
   \hspace{-0.2in}
   \subfigure{
    \includegraphics[width=1.4in]{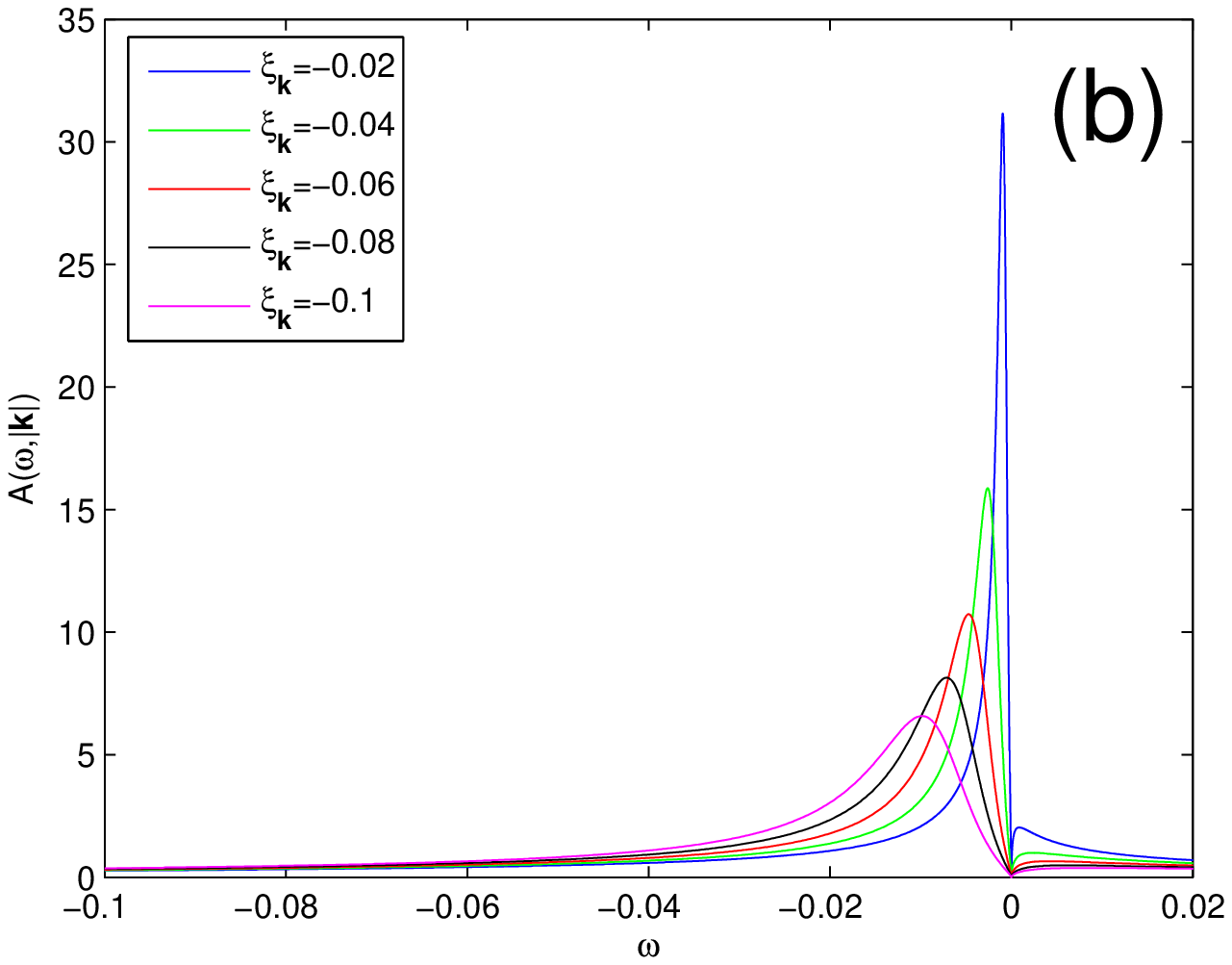}}
    \\
    \vspace{-0.15in}
   \subfigure{
    \includegraphics[width=1.4in]{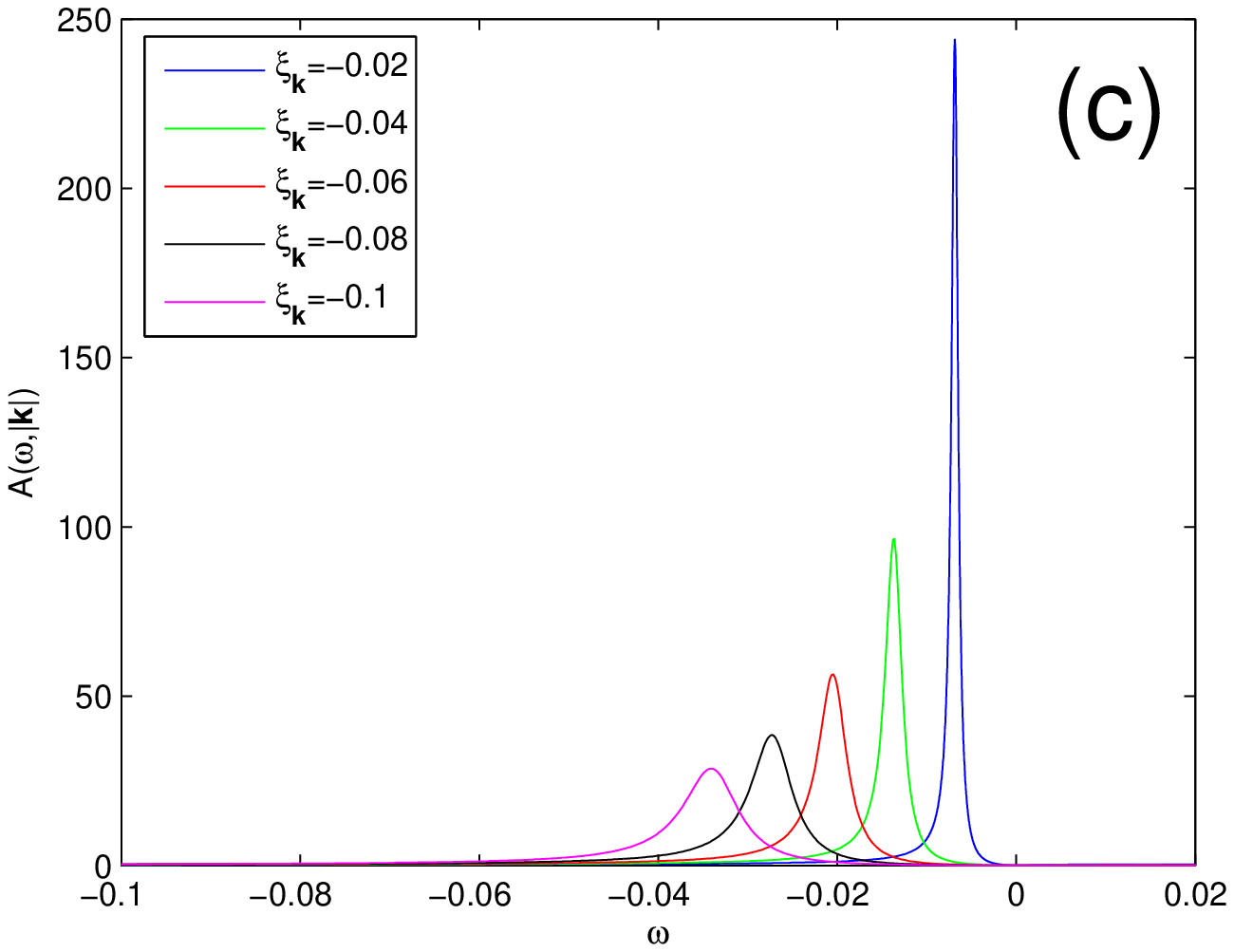}}
   \hspace{-0.2in}
   \subfigure{
    \includegraphics[width=1.4in]{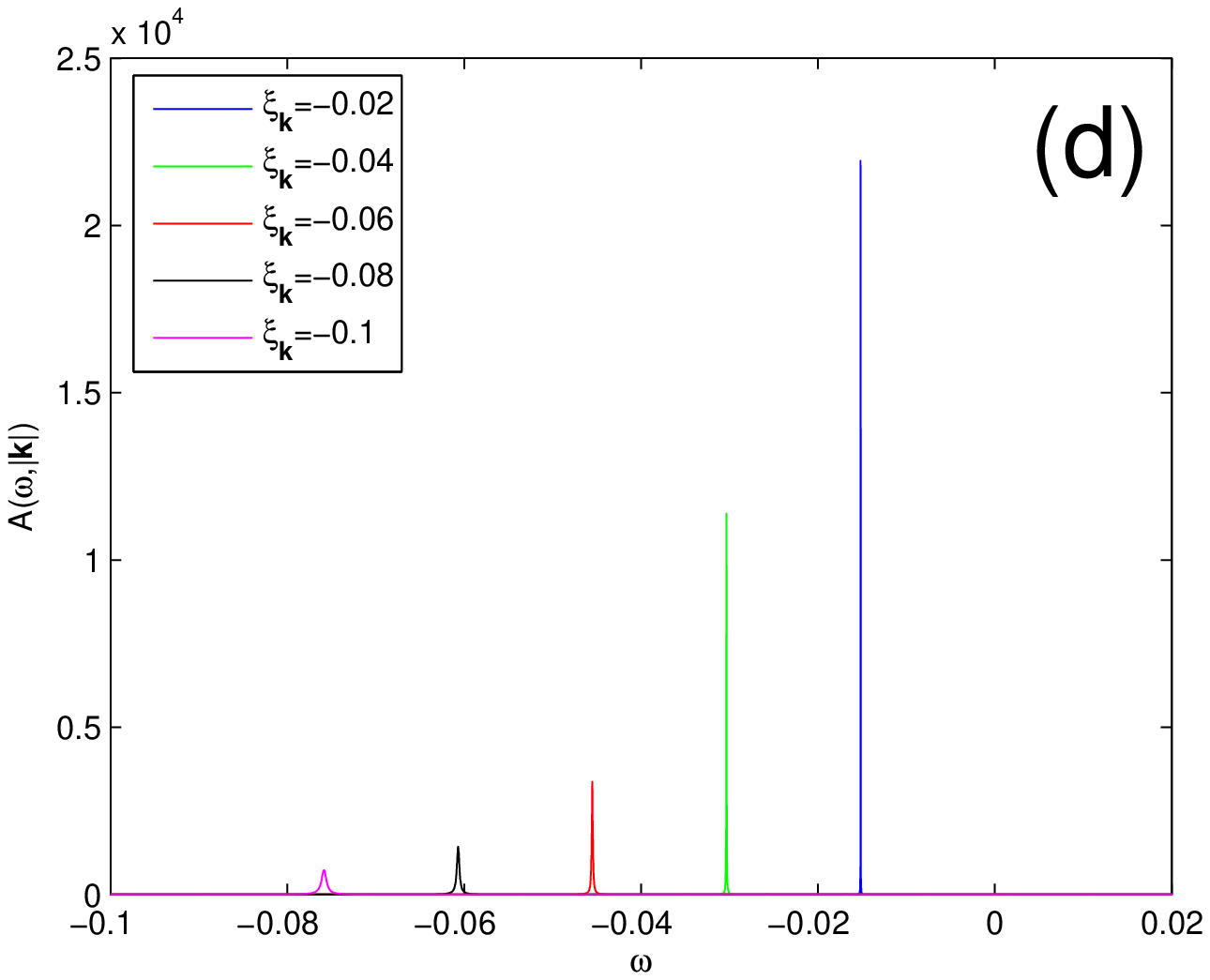}}
   \vspace{-0.2in}
\caption{Spectral function. (a) $0<x<\frac{1}{2}$; (b)
$\frac{1}{2}\leq x\leq1$; (c) $1<x\leq2$; (d) $x>2$.}
     \label{fig: EDC}
\end{figure}

Using these self-energy functions, it is easy to obtain that $Z=0$
for $0<x\leq 1$ and $0<Z<1$ for $x>1$. From the spectral functions
at various $x$ shown in Fig.\ref{fig: EDC}, we see that the
well-defined quasiparticle peak exists only for $x>1$ where $Z\neq
0$. The residue $Z$ also represents the discontinuity of momentum
occupation number $n(\mathbf{k})$ at $|\mathbf{k}| = k_F$. When
$0<x\leq 1$, there is no discontinuity in $n(\mathbf{k})$; but when
$x>1$, there is a discontinuity in $n(\mathbf{k})$. In conventional
many-body theory, $Z$ can uniquely differentiate a non-FL from the
normal FL. In strongly correlated systems, however, $Z$ itself can
no longer fully characterize the non-FL. In order to see how this
happens, we will calculate $n(\mathbf{k})$ and its derivative for
all values of $x$.

For $0<x<1$, the momentum occupation number is
\begin{eqnarray}
n(\mathbf{k}) \propto \int_{-\infty}^{0}d\omega
\frac{|\omega|^x}{\left[\omega - \xi_{\mathbf{k}} + C
I(x)|\omega|^x\right]^2 +\left[C|\omega|^x\right]^2}. \nonumber
\end{eqnarray}
At Fermi energy, its derivative with respect to $|\mathbf{k}|$ is
\begin{eqnarray}
\left.n^{\prime}(\mathbf{k}) \right|_{k_F} \propto
\int_{-\infty}^{0} d\omega \frac{|\omega|^x \left(\omega + CI(x)
|\omega|^x\right)}{\left[\left(\omega+CI(x)|\omega|^x\right)^2
+\left(C|\omega|^x\right)^2\right]^2}. \nonumber
\end{eqnarray}
As $\omega \rightarrow 0$, the integrand approaches
$-\frac{1}{|\omega|^{2x}}$. For $0<x<\frac{1}{2}$, it is easy to
obtain
\begin{equation}
\left.n^{\prime}(\mathbf{k})\right|_{k_F-0} = \mbox{finite},\qquad
\left.n^{\prime}(\mathbf{k})\right|_{k_F+0} = \mbox{finite}.
\end{equation}
For $\frac{1}{2}\leq x<1$, $n^{\prime}(\mathbf{k})$ diverges as:
\begin{equation}
\left.n^{\prime}(\mathbf{k})\right|_{k_F-0} \rightarrow
-\infty,\qquad \left.n^{\prime}(\mathbf{k})\right|_{k_F+0}
\rightarrow -\infty,
\end{equation}
which indicates that the momentum occupation number drops
dramatically at $k_F$.

Similarly, for $x=1$, we can obtain the following derivative of
occupation number
\begin{eqnarray}
\left.n^{\prime}(\mathbf{k})\right| _{k_F} \propto
\int_{-\omega_c}^{0} d\omega \frac{|\omega|\left(\omega -
C\frac{2}{\pi}\omega\ln\left(\frac{\omega_c}{|\omega|}\right)\right)
}{\left[\left(\omega-C\frac{2}{\pi}\omega
\ln\left(\frac{\omega_c}{|\omega|}\right)\right)^2
+\left(C|\omega|\right)^2\right]^2}. \nonumber
\end{eqnarray}
As $\omega \rightarrow 0$, the integrand approaches
$-\frac{1}{|\omega|^2
\left[\ln\left(\frac{\omega_{c}}{|\omega|}\right)\right]^3}$, thus
\begin{equation}
\left.n^{\prime}(\mathbf{k})\right|_{k_F-0} \rightarrow
-\infty,\qquad \left.n^{\prime}(\mathbf{k})\right|_{k_F+0}
\rightarrow -\infty,
\end{equation}
which is similar to the case of $\frac{1}{2}\leq x<1$.

For $x>1$, we have
\begin{eqnarray}
\left.n^{\prime}(\mathbf{k})\right| _{k_F} \propto
\int_{-\omega_{c}}^{0}d\omega \frac{|\omega|^{x}
\left(\omega-C\frac{2}{\pi}\frac{|\omega_c|^{x-1}}{x-1}
\omega\right)}{\left[\left(\omega -
C\frac{2}{\pi}\frac{|\omega_c|^{x-1}}{x-1}\omega\right)^2 +
\left(C|\omega|^{x}\right)^2\right]^2}. \nonumber
\end{eqnarray}
As $\omega \rightarrow 0$, the integrand approaches
$-\frac{1}{\left|\omega\right|^{3-x}}$. For $1<x\leq 2$, the residue
$Z\neq 0$ and the derivative of $n(\mathbf{k})$ is
\begin{equation}
\left.n^{\prime}(\mathbf{k})\right|_{k_F-0} \rightarrow
-\infty,\qquad \left.n^{\prime}(\mathbf{k})\right|_{k_F+0}
\rightarrow -\infty.
\end{equation}
For $x>2$, although $Z\neq 0$, we know that
\begin{equation}
\left.n^{\prime}(\mathbf{k})\right|_{k_F-0} = \mbox{finite},\qquad
\left.n^{\prime}(\mathbf{k})\right|_{k_F+0} = \mbox{finite}.
\end{equation}

\begin{figure}[h]
   \subfigure{
    \includegraphics[width=1.3in]{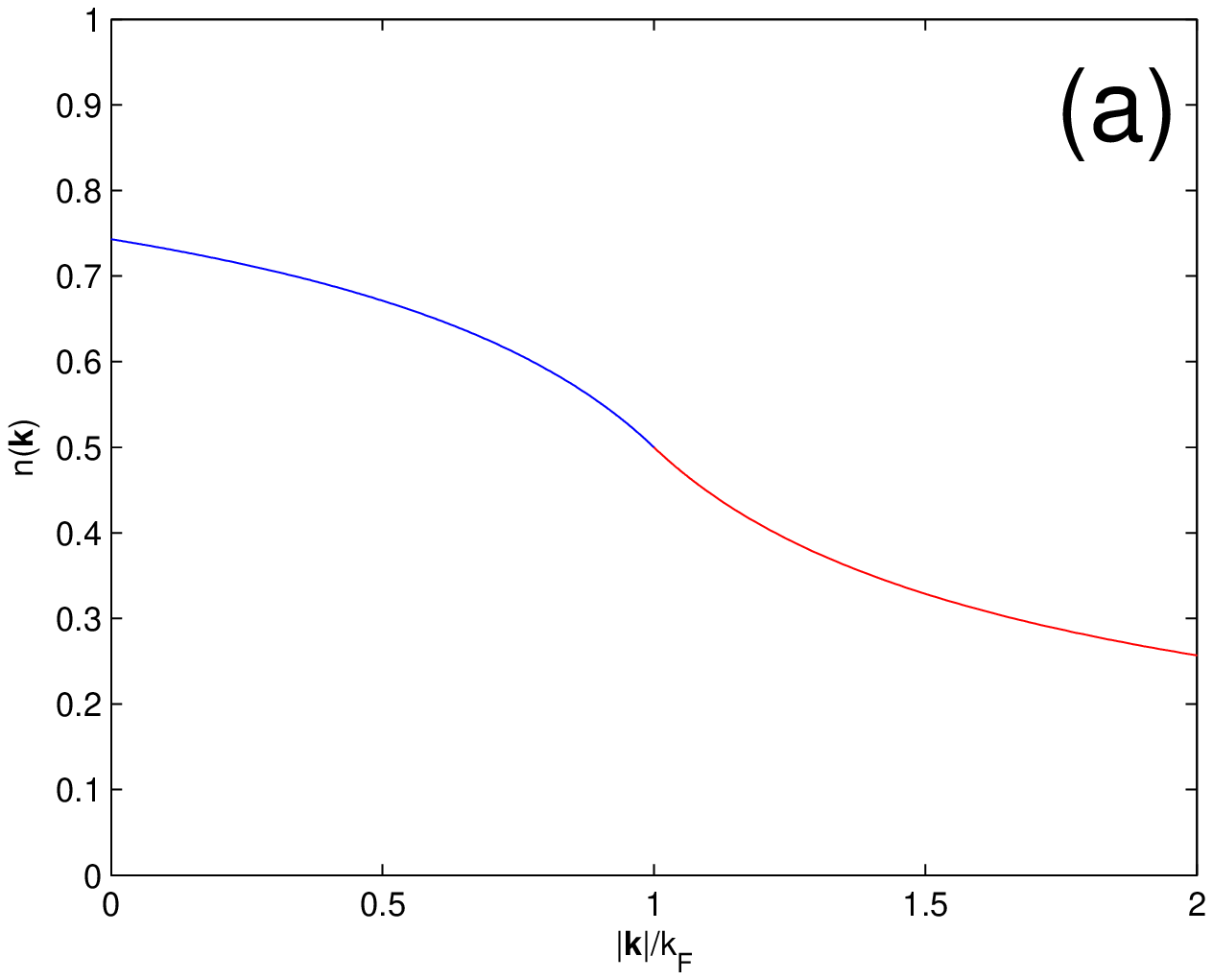}}
   \hspace{-0.2in}
   \subfigure{
    \includegraphics[width=1.3in]{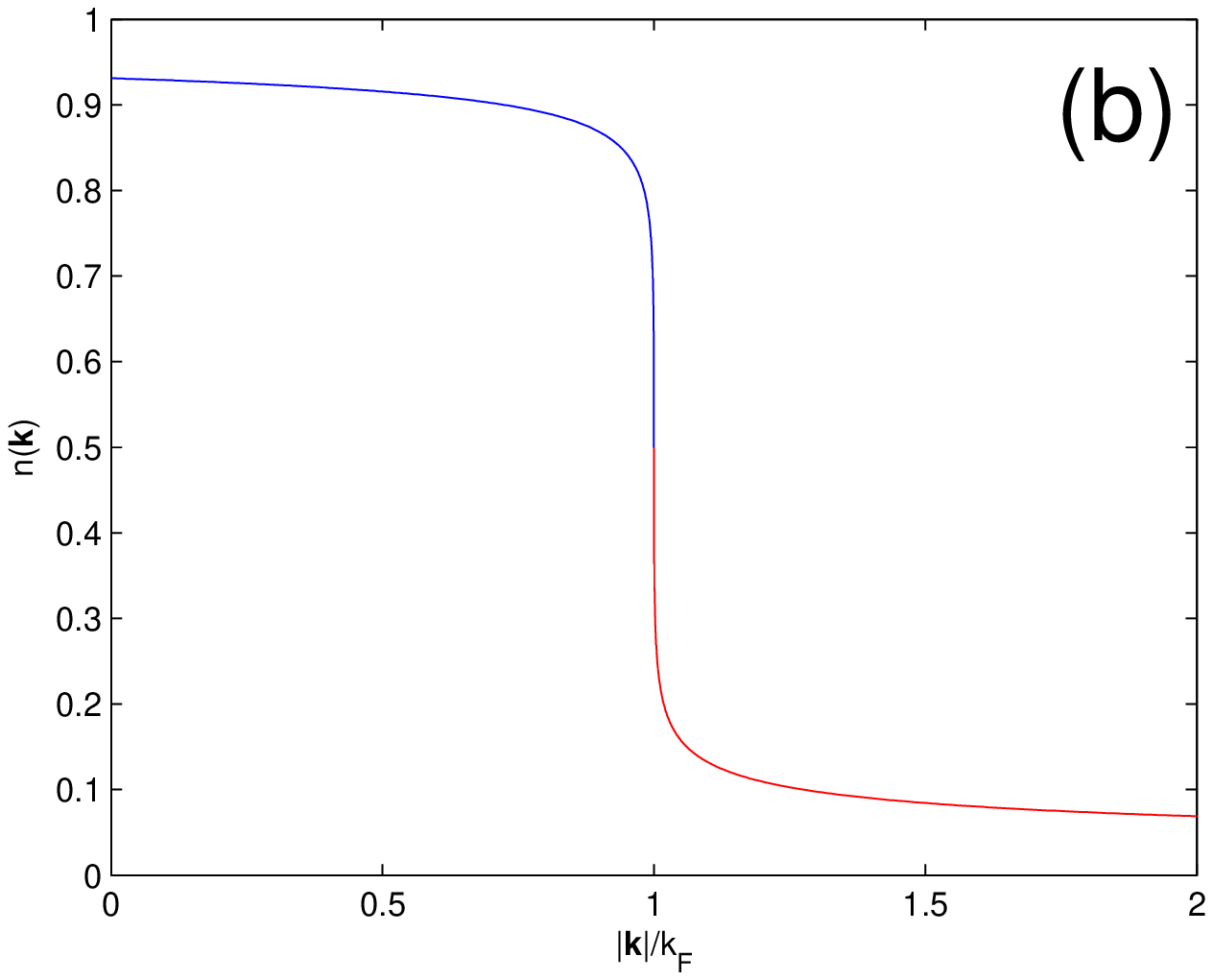}}
    \\
   \vspace{-0.15in}
   \subfigure{
    \includegraphics[width=1.3in]{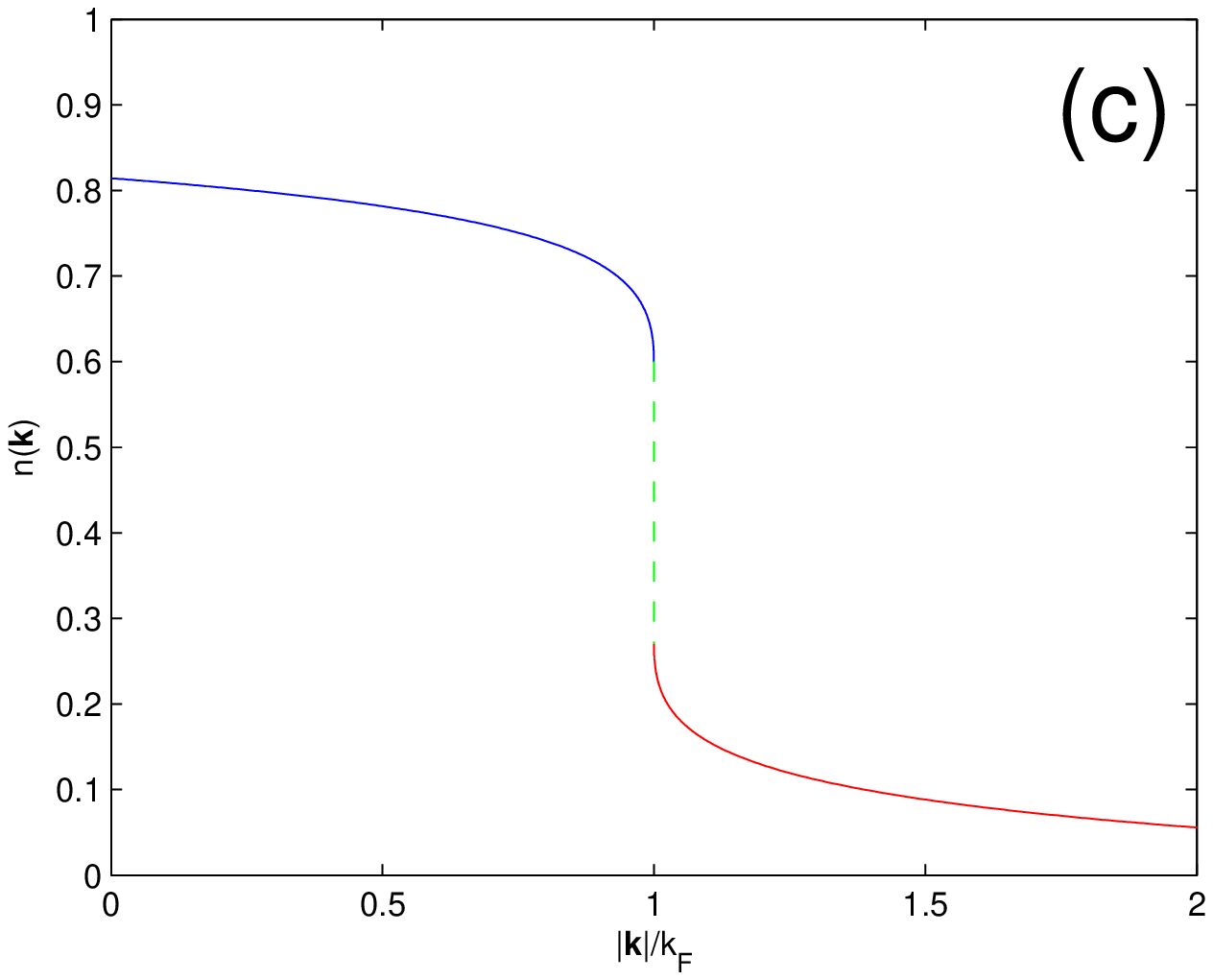}}
   \hspace{-0.2in}
   \subfigure{
    \includegraphics[width=1.3in]{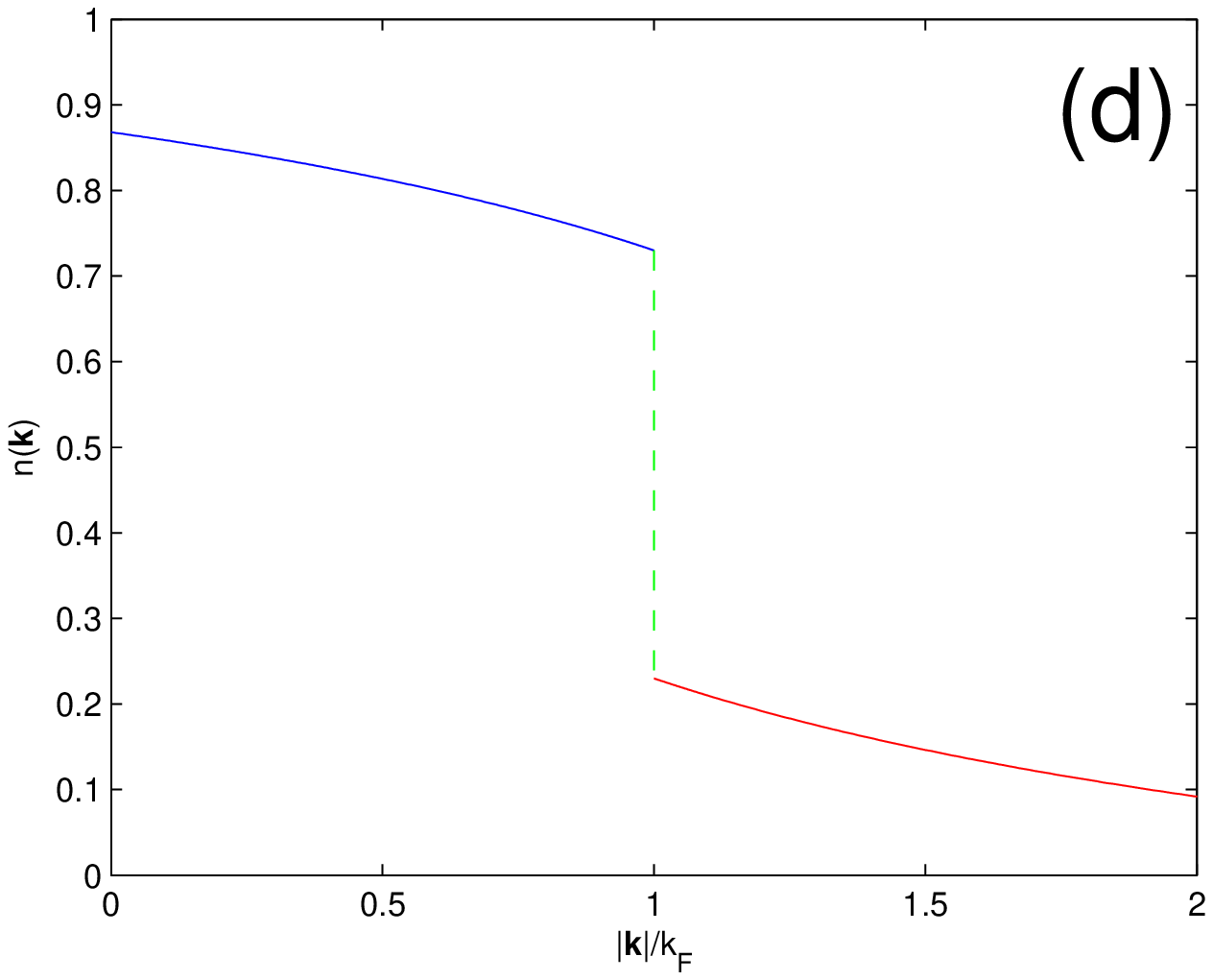}}
   \vspace{-0.2in}
\caption{Fermion momentum occupation number $n(\mathbf{k})$. (a)
$0<x<\frac{1}{2}$; (b) $\frac{1}{2}\leq x\leq 1$; (c) $1<x\leq 2$;
(d) $x>2$.}
     \label{fig: FermiSurface}
\end{figure}

\begin{figure}[h]
   \subfigure{
    \includegraphics[width=1.3in]{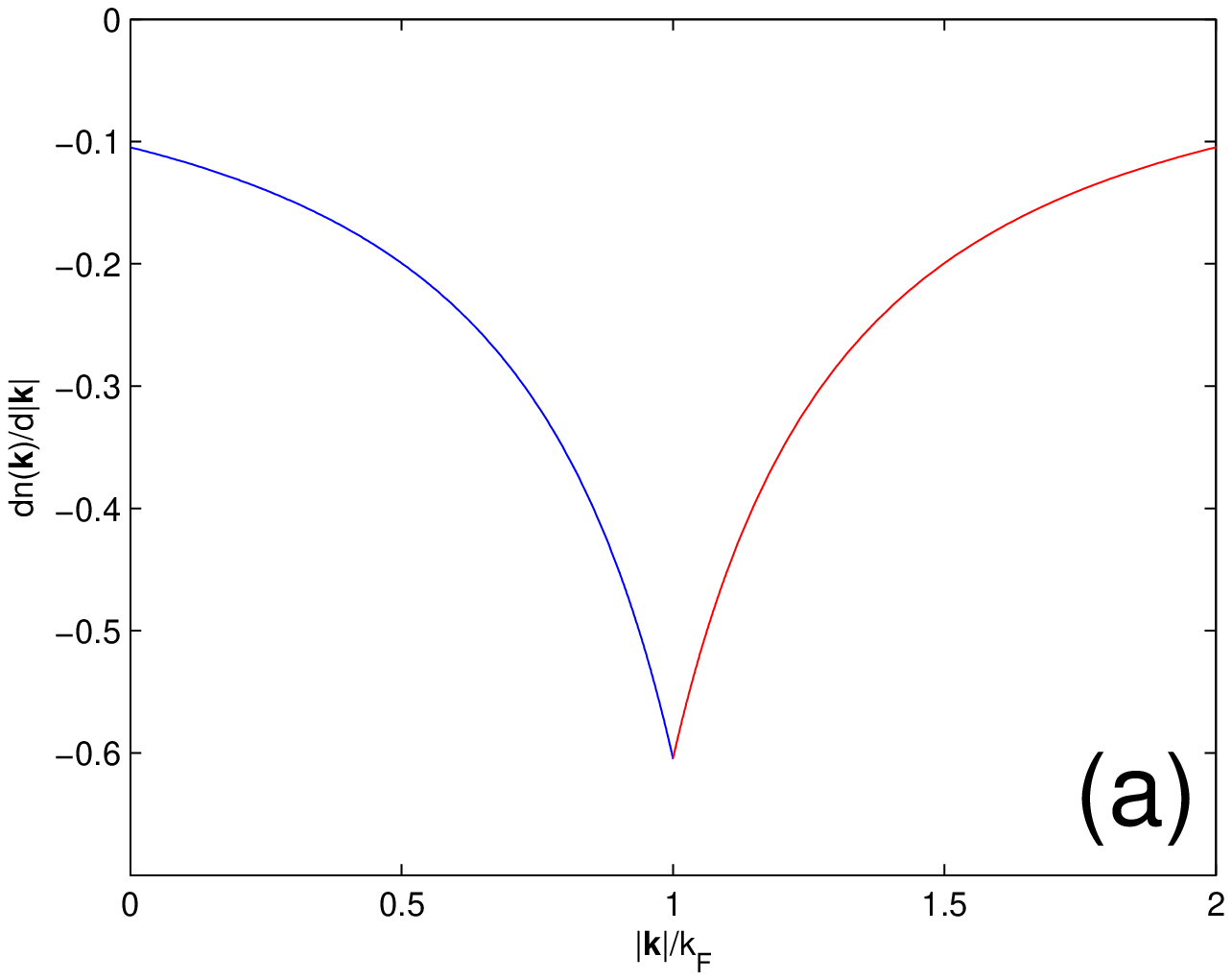}}
   \hspace{-0.2in}
   \subfigure{
    \includegraphics[width=1.3in]{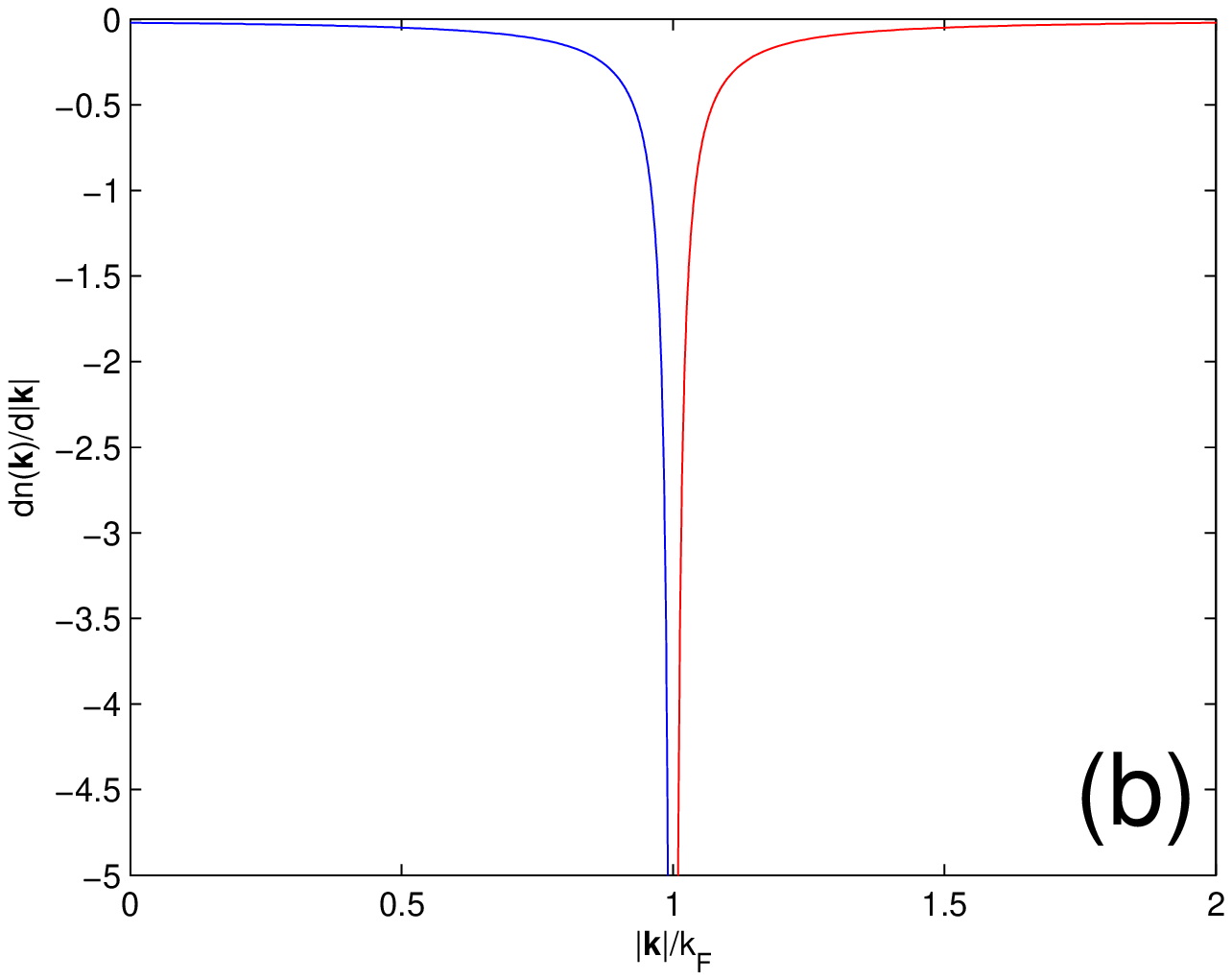}}
   \\
   \vspace{-0.15in}
   \subfigure{
    \includegraphics[width=1.3in]{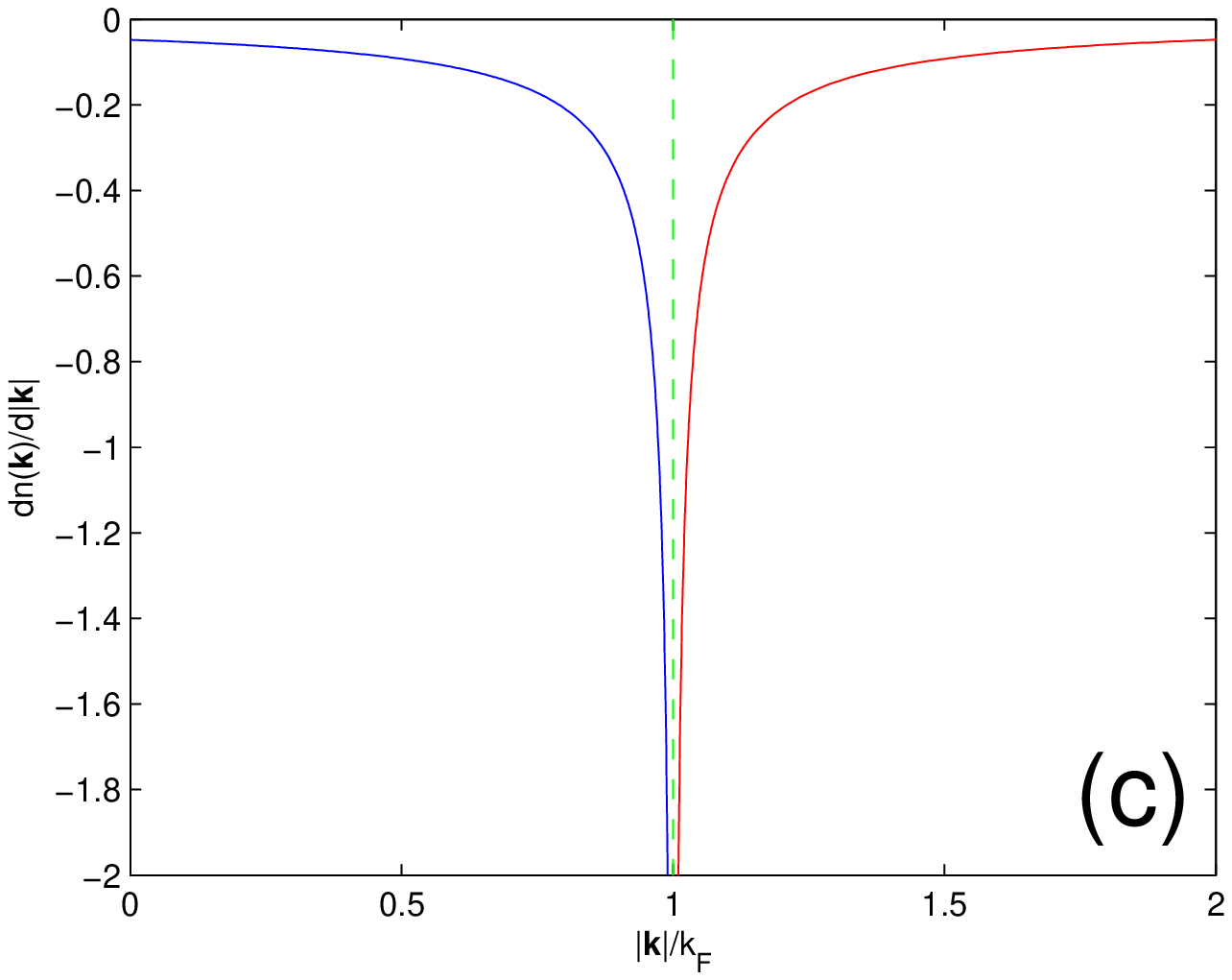}}
   \hspace{-0.2in}
   \subfigure{
    \includegraphics[width=1.3in]{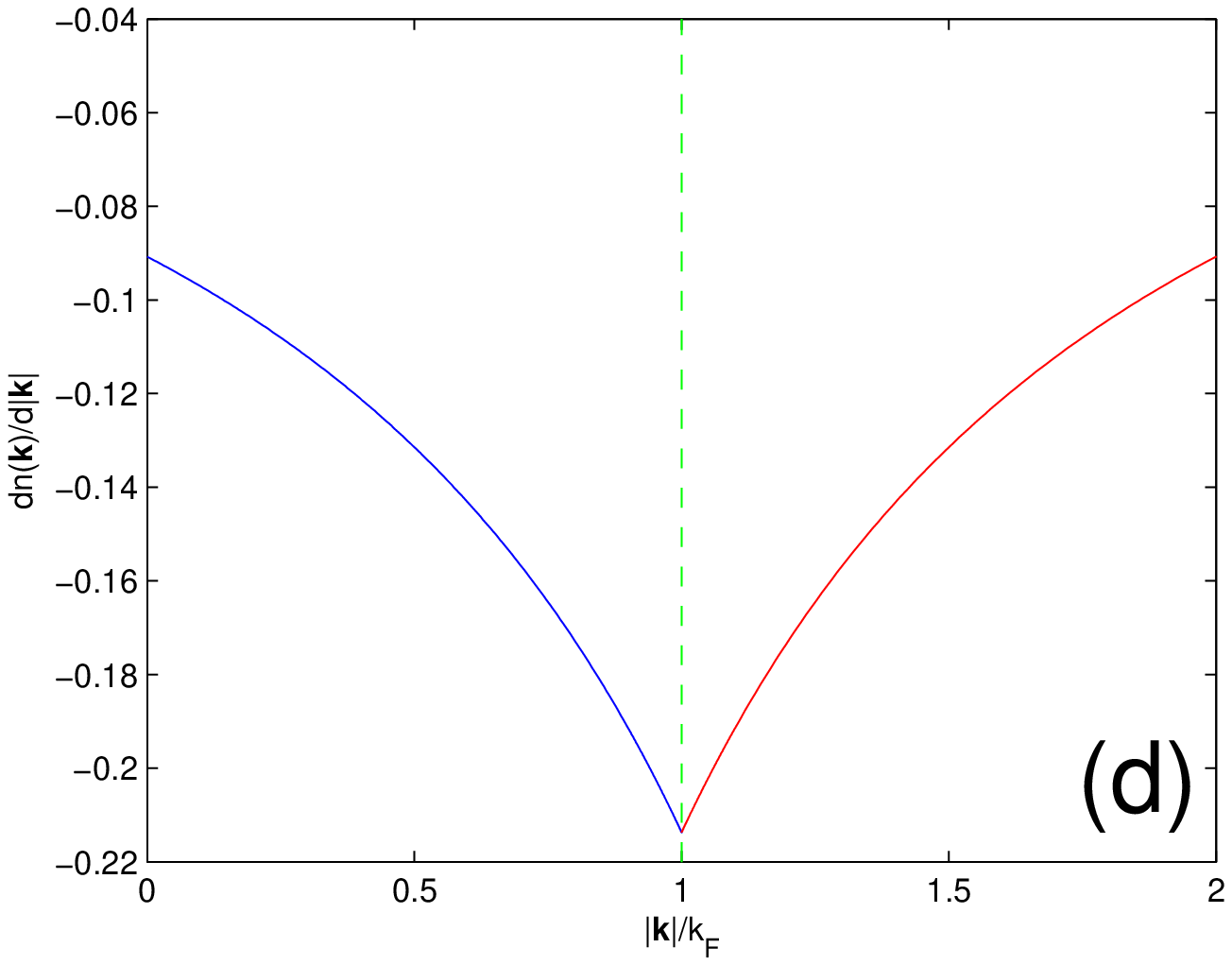}}
   \vspace{-0.2in}
\caption{Derivative of $n(\mathbf{k})$. (a) $0<x<\frac{1}{2}$; (b)
$\frac{1}{2}\leq x\leq1$; (c) $1<x\leq2$; (d) $x>2$.}
     \label{fig: DerivativeNK}
\end{figure}

The dependence of momentum occupation number $n(\mathbf{k})$ on $x$
is shown in Fig.\ref{fig: FermiSurface}. Apparently, as the
parameter $x$ grows, the interacting system of massless Dirac
fermions first develops a sharp Fermi surface at $x = \frac{1}{2}$
and then develops a finite quasiparticle residue $Z$ once $x$
exceeds unity. There is a fundamental difference between the
$n(\mathbf{k})$ functions for $0<x<\frac{1}{2}$ and for
$\frac{1}{2}\leq x\leq 1$. In the former case, the residue $Z=0$, so
there are no well-defined quasiparticles. Moreover, the derivative
of $n(\mathbf{k})$ is continuous at the Fermi energy $k_F$, so there
is also no Fermi surface. In the latter case, the residue $Z=0$ and
thus there is no well-defined quasiparticle peak. However, the
derivative of $n(\mathbf{k})$ diverges at $k_F$, therefore
$n(\mathbf{k})$ drops suddenly as $|\mathbf{k}|$ increases across
$k_F$, which can be identified as the presence of a sharp Fermi
surface. To see more details of the evolution of $n(\mathbf{k})$
with $x$, we also show the derivative of $n(\mathbf{k})$ in
Fig.\ref{fig: DerivativeNK}.

\begin{figure}[h]
    \subfigure{
    \label{fig:NK0}
   \includegraphics[width=1.3in]{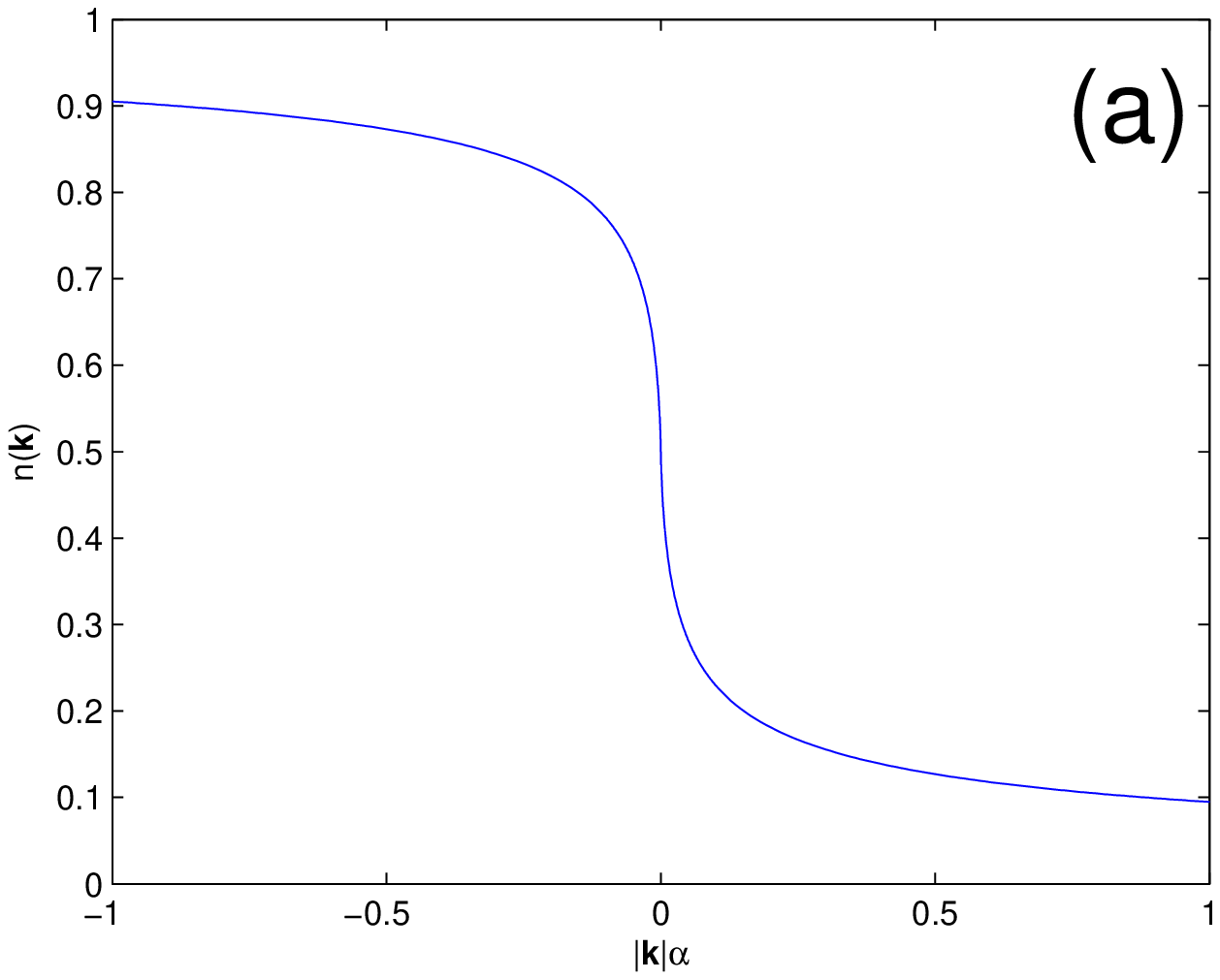}}
   \hspace{-0.2in}
    \subfigure{
    \label{fig:NK1}
    \includegraphics[width=1.3in]{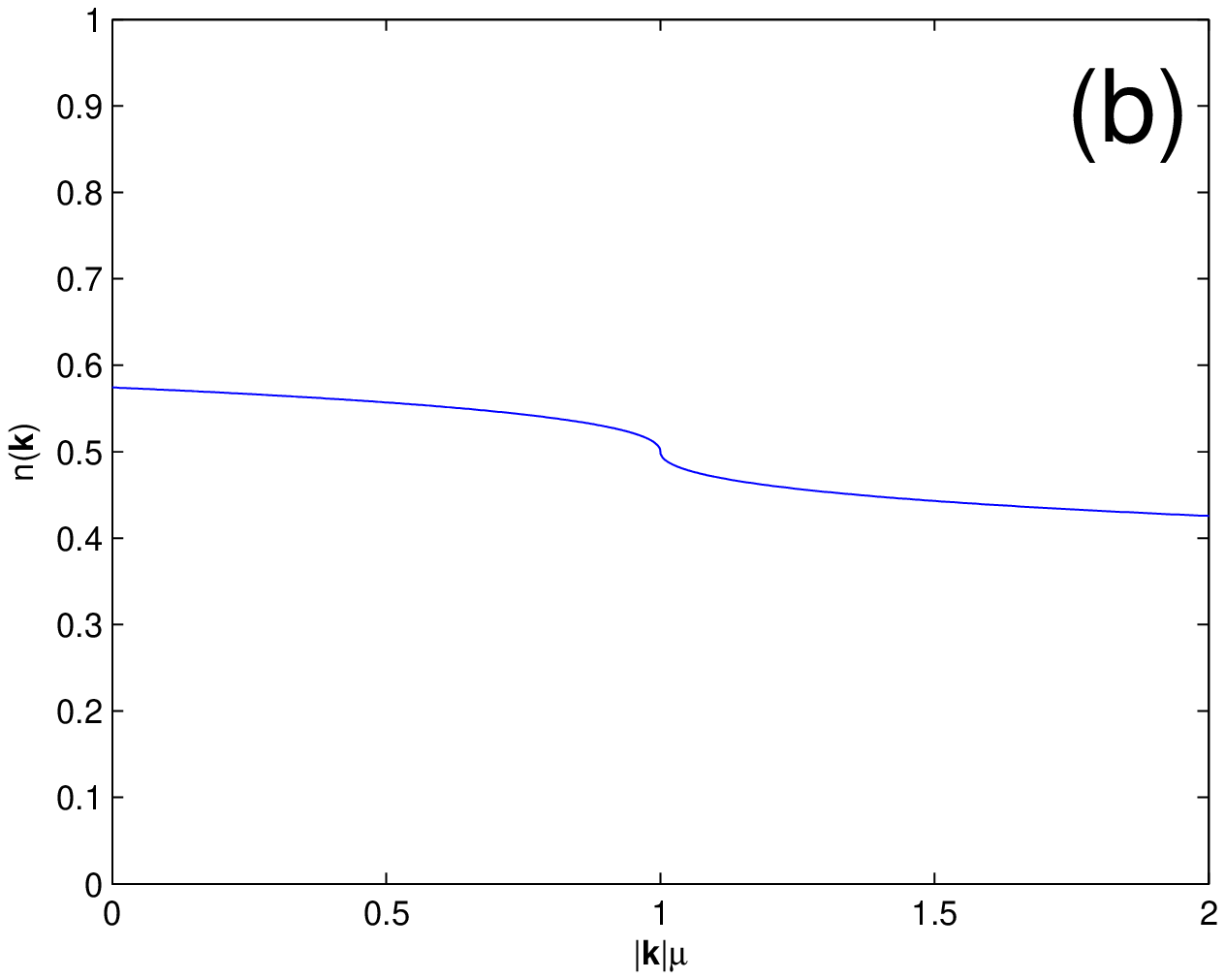}}
    \\
    \vspace{-0.15in}
    \subfigure{
    \label{fig:NK2}
    \includegraphics[width=1.3in]{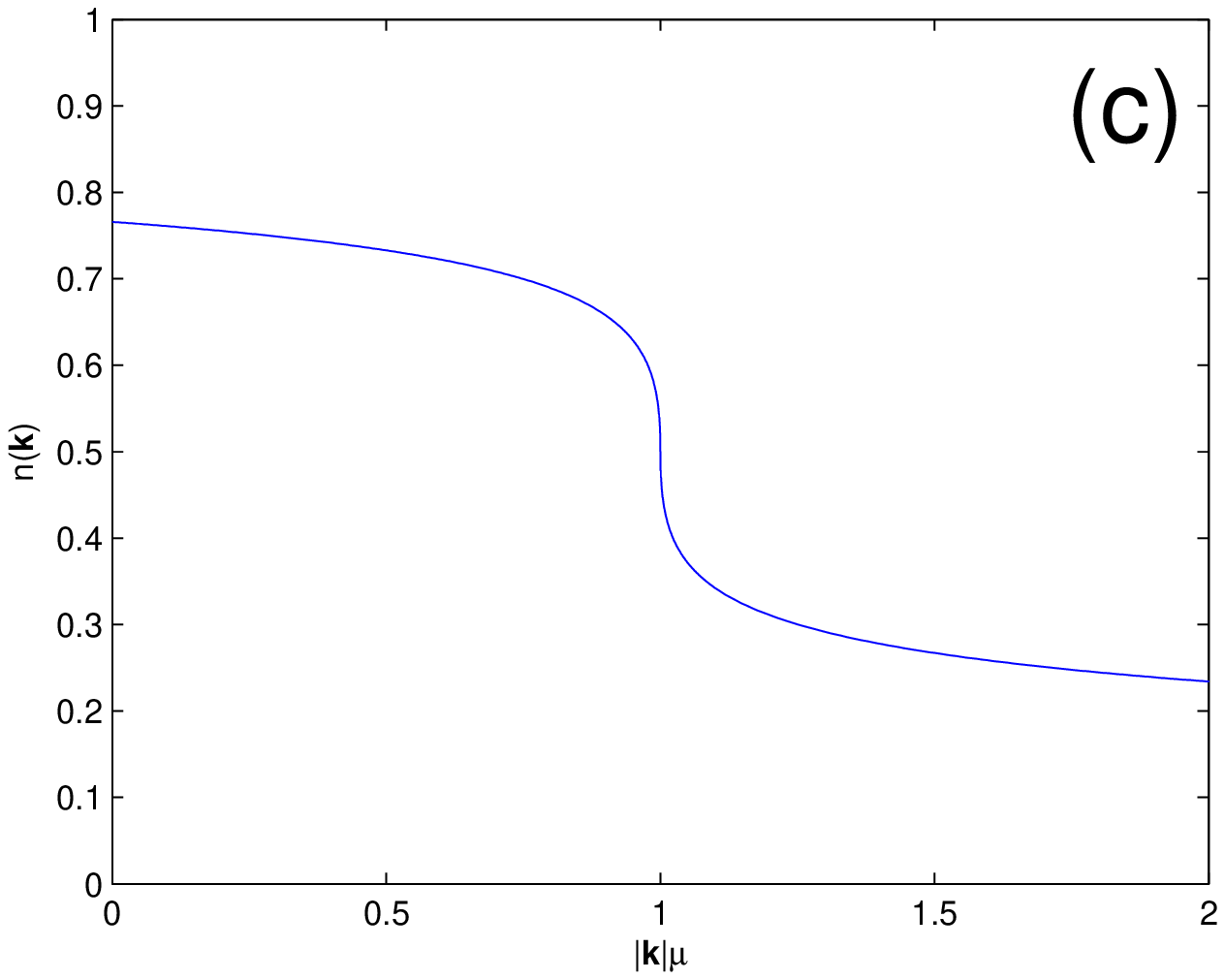}}
    \hspace{-0.2in}
     \subfigure{
    \label{fig:NK3}
    \includegraphics[width=1.3in]{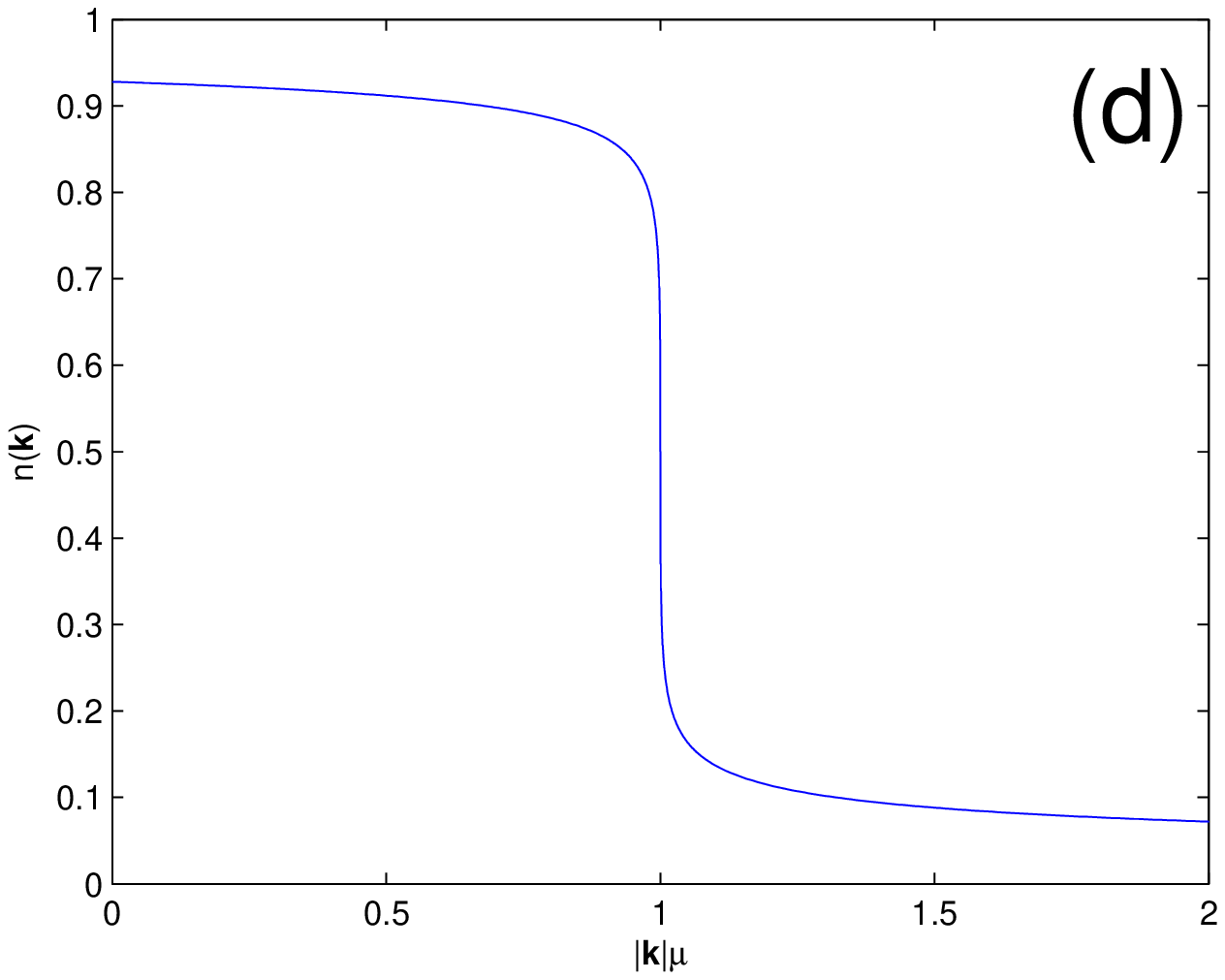}}
    \vspace{-0.2in}
\caption{$n(\mathbf{k})$ for different chemical potential $\mu$ in
QED$_3$. $\mu$ is zero in (a) and increases from (b) to (d).}
     \label{fig:NK}
\end{figure}

In summary, we can divide all the interacting Dirac fermion systems
into three classes: I) for $0<x<\frac{1}{2}$, there are no sharp
Fermi surface and no well-defined quasiparticle peak; II) for
$\frac{1}{2}\leq x\leq 1$, there is sharp Fermi surface but no
well-defined quasiparticle peak; III) for $x>1$, there are both
sharp Fermi surface and well-defined quasiparticle peak.

\begin{figure}[h]
   \subfigure{
    \includegraphics[width=1.3in]{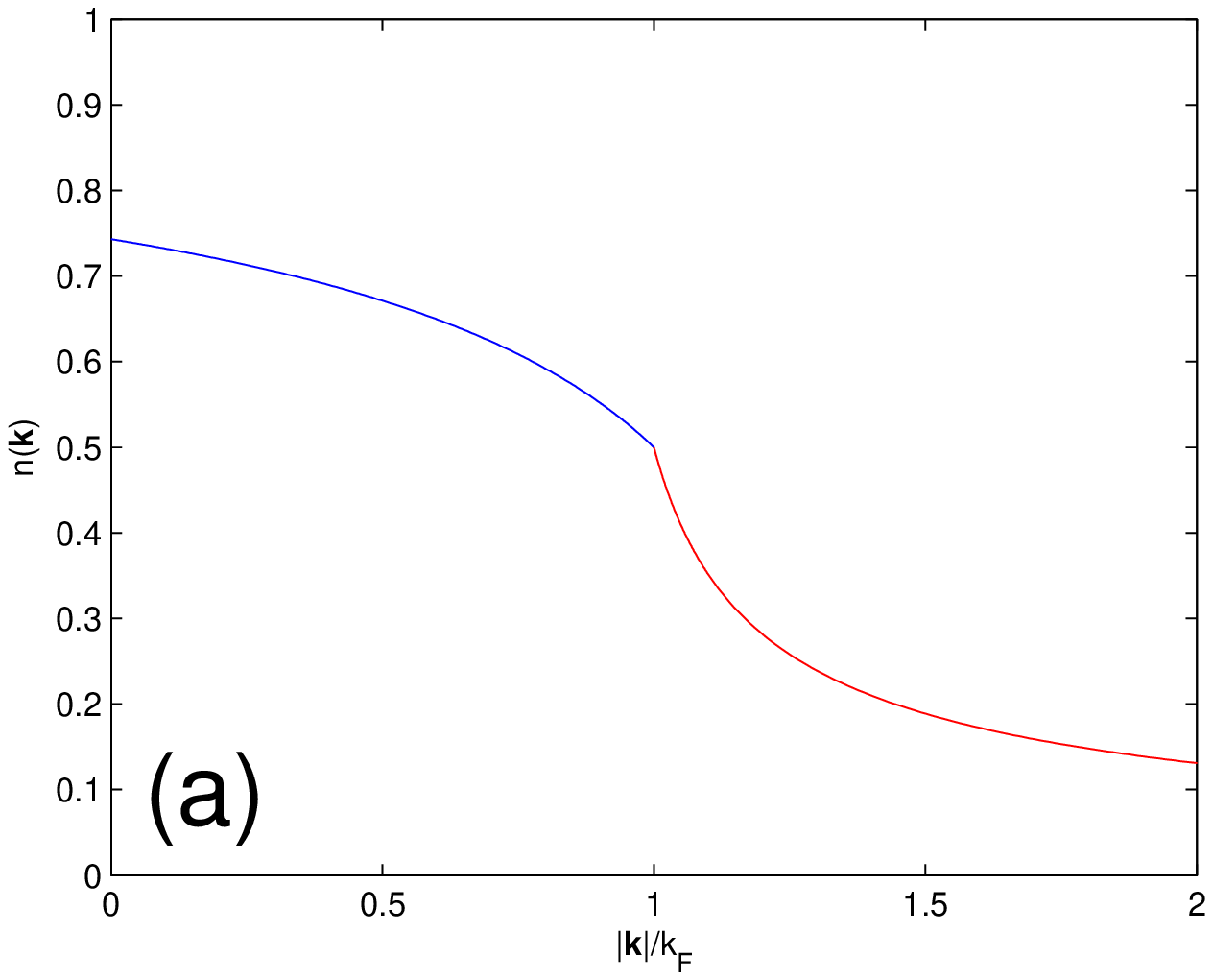}}
   \hspace{-0.2in}
   \subfigure{
    \includegraphics[width=1.3in]{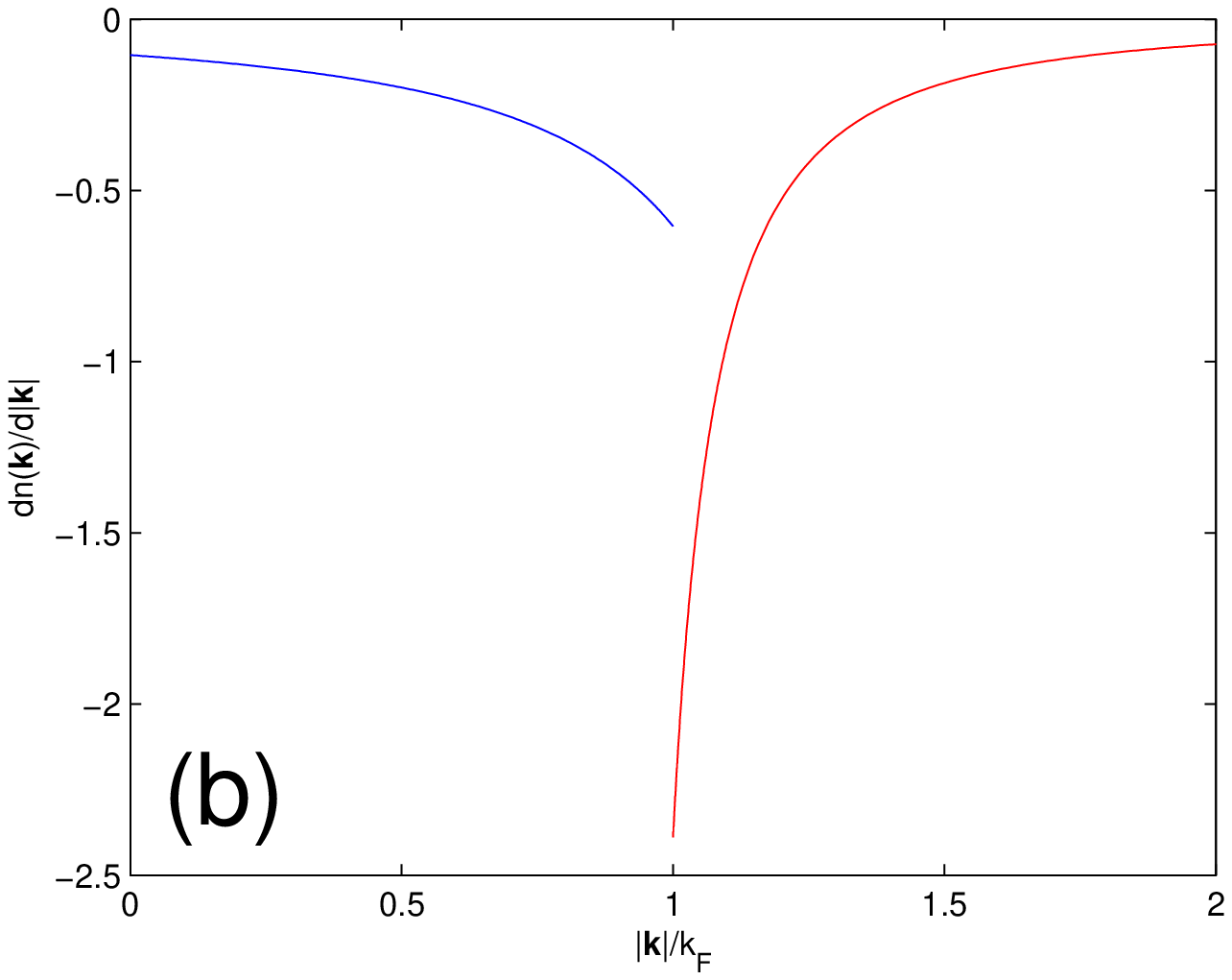}}
   \vspace{-0.2in}
\caption{Critical Fermi surface of Senthil. (a) $n(\mathbf{k})$; (b)
$n^{\prime}(\mathbf{k})$.}
     \label{fig: SenthilCFS}
\end{figure}

In QED$_3$, the fermion decay rate has $x=\frac{1}{2}$ at zero $\mu$
and $x=\frac{2}{3}$ at finite $\mu$. These states fall into the
class II. To see the Fermi surface evolution with growing $\mu$, we
show $n(\mathbf{k})$ in Fig.\ref{fig:NK}. We can see that
$n(\mathbf{k})$ does not evolve continuously from $\mu=0$ to the
case of finite $\mu$. On the contrary, as $\mu$ jumps from zero to a
small value, the curve of $n(\mathbf{k})$ changes dramatically.
However, the Fermi surface remains sharply defined for arbitrary
small value of $\mu$, because $\mu$ does not alter the energy
dependence of decay rate \cite{WangLiu2}. The marginal FL ($x=1$)
caused by the long-range Coulomb interaction or certain massless
order parameter fluctuation also belongs to class II. In the
presence of short-ranged gauge/Coulomb interaction, the decay rate
has exponent $x > 1$ and thus the behavior falls into class III.
Unfortunately, it is currently unclear what kind of interactions can
produce behaviors belonging to class I, which deserves further
research.

We emphasize that the unconventional Fermi surface studied in this
paper is different from that of Senthil \cite{Senthil}. As shown in
Fig.\ref{fig: SenthilCFS}, the critical Fermi surface proposed by
Senthil is defined by the discontinuity in the derivative of
$n(\mathbf{k})$. In our problem, however, there is no such kink
singularity and the sharp Fermi surface is characterized by the
divergence of the derivative of $n(\mathbf{k})$.

Finally, although all the above discussions were made in the case of
massless Dirac fermions, the unconventional Fermi surface of class
II should also be realized in a number of non-relativistic strongly
correlated systems, such as non-relativistic fermion-gauge system
with large Fermi surface \cite{Lee92} and quantum critical metals
near various symmetry-breaking instabilities \cite{Metzner, Mross},
provided that the exponent in fermion decay rate satisfies
$\frac{1}{2}\leq x\leq 1$.

This work was supported by the National Science Foundation of China
under Grant No. 10674122.

\end{document}